# Vibrational two-photon emission from coherently excited solid parahydrogen.

Yuki Miyamoto[a)], Hideaki Hara, Takahiko Masuda, Noboru Sasao, Satoshi Uetake, Akihiro Yoshimi, Koji Yoshimura, and Motohiko Yoshimura

*Research Institute for Interdisciplinary Science, Okayama University, Okayama 700-8530, Japan.*

We report observation of two-photon emission from coherently excited vibrational state of solid parahydrogen, known as one of quantum solids. The coherent state between the ground and the excited states is prepared by stimulated Raman scattering using two visible laser pulses. The two-photon emission is triggered by another mid-infrared laser pulse. The observed two-photon emission persists even when the trigger pulse is injected long after the excitation. The observed phenomenon is due to a long decoherence time of the vibrational states of solid parahydrogen, which is attributed to special nature as quantum solid. Dependence of the decoherence on target temperature and residual orthohydrogen concentration are studied along with its time evolution. It is found that the emission intensity increases even after the excitation pulses pass through the target completely. The coherence development is highly suppressed at high target temperature and high residual orthohydrogen concentration. Effects of target annealing and laser-induced damage on the target are also observed.

**I. INTRODUCTION**

Two-photon absorption and Raman scattering are ubiquitous two-photon processes in chemistry, spectroscopy, molecular physics, quantum optics, and their related fields. Two-photon emission (TPE), a reversal process of the two-photon absorption, is less common than the above two processes. Nevertheless, TPE is known to be an important process especially in astrochemistry, astrophysics and atomic physics [1-3] and has been studied experimentally and theoretically for decades, mainly in gaseous phase. On the other hand, there are only a limited number of reports on observation of TPE in condensed phases.[4-10] These stimulated and spontaneous TPEs were observed in semiconductors. As far as we know, there is no report on TPE in other types of solids. Several applications of TPE in solid are proposed in quantum optics and quantum information.[11-17]

The difficulty of observing TPE lies in its weakness. Transition intensity of TPE is a few orders of magnitude smaller than that of single photon process. One of dominant competing processes is single photon emission. Non-radiative processes are other competing processes especially in solids where they are quite fast. We have adopted coherent amplification in order to overcome them. In our previous papers,[18, 19] we reported TPE from the vibrational states of gaseous parahydrogen excited by the Raman scattering. It was found that TPE was enhanced by the coherence and its intensity was increased more than $10^{18}$ times larger than that of spontaneous TPE rate.

The coherent amplification is amplification by constructive interference between coherently excited particles. A famous example of coherent amplifications is the superradiance.[20] The present coherent TPE is a two-photon version of the

---

[a)] Electronic mail: miyamo-y@cc.okayama-u.ac.jp

superradiance so to say. Coherent amplification can be a powerful method in high density targets like solids because its gain is proportional to the square of excitation density. However, the coherence is generally more fragile in solids than that in gas phases due to strong interactions with environment. It is important to prepare stable coherent states in order to observe TPE using coherent amplification. It is also useful for observation of other coherent phenomena in solids.

The vibrational state of solid parahydrogen is an ideal system for observation of TPE using the coherent amplification. First of all, vibrational transition of homonuclear diatomic molecules is a single-photon forbidden and two-photon allowed two-level system so that there is effectively no undesired single-photon processes. Secondly, the vibrational states of solid parahydrogen have long decoherence time in comparison with general solids by virtue of its extraordinary nature as quantum solid.[21, 22] The vibrational states of solid parahydrogen have been studied well.[23-29] The linewidth of its vibrational Raman transition to the $v = 1$, $J = 0$ state was found to be extremely narrow.[25-27] The decoherence time is estimated to be a few tens of nanosecond from the linewidth, which is consistent with the result of coherent anti-Stokes Raman spectroscopy (CARS).[28] It is several orders of magnitude longer than those in typical solids.

In the present paper, we report the observation of TPE from the coherently excited $v = 1$ state of solid parahydrogen. The coherent state is prepared by the stimulated Raman process with two visible laser pulses. The two-photon emission is triggered by another mid-infrared pulse. Dependences of TPE intensity on trigger timing, target temperature and residual impurity concentration are measured. The process reported here is not spontaneous emission but triggered by external field so that one may consider the process as common four-wave mixing or parametric generation. However, resonance with the real vibrational state and the persistent coherence of solid parahydrogen make situation quite different as discussed later. The present study shows that solid parahydrogen provides ideal environment to study coherent phenomena in solid states. The present result also provides new types of frequency conversion techniques and two-photon sources.

The authors have proposed the methodology to investigate the unrevealed properties of neutrinos, such as their absolute masses, by observing coherently amplified emission of photon and neutrinos from excited atoms or molecules (neutrino mass spectroscopy).[30] Observation of TPE enhanced by the coherence is the first step to the neutrino mass spectroscopy because it proves usefulness of the coherent amplification to observe rare mechanisms involving plural particles.[18, 19] Observation of the coherent amplification in solids is also an important step for the neutrino mass spectroscopy because it opens the possibility of higher enhancement.

The rest of the paper is organized as follows. In the next section, experimental setup and methods: target preparation, energy scheme, laser systems, and detection system, are described. In the results and discussions section, an observed TPE spectrum and dependence of the TPE intensity on several experimental conditions are reported and compared with previous works. The last section is devoted to the conclusions.

## II. EXPERIMENTAL SETUP

### A. Preparation of solid parahydrogen.

Preparation of solid parahydrogen target is similar to that described in our previous paper.[29] Normal hydrogen gas is converted to parahydrogen gas by magnetic catalyst cooled to about 14 K. Prepared parahydrogen gas is flowed into a copper cell of 5 mm in length and 20 mm in diameter. Both ends of the cell are sealed with barium-fluoride windows and indium wires. Temperature of the cell is kept to about 8 K during crystal growth. Then, the cell is cooled to about 4 K. Annealing of the crystal changes the behavior of TPE as discussed in the next section. The annealing is performed by increasing the crystal temperature above 10 K for 30 minutes. We confirm that annealing longer than 30 minutes does not affect experimental results largely. Target temperature during measurement is about 4.1 K, if not otherwise described. Parahydrogen sample contains small amount of residual orthohydrogen. Its concentration is estimated by measuring infrared absorption of solid parahydrogen induced by the residual orthohydrogen. It is known that the absorption intensity is proportional to orthohydrogen concentration in the low concentration limit.[31, 32] The spectra are measured by a Fourier-transform infrared spectrometer (Nicholet 6700 FT-IR) with resolution of 0.125 cm$^{-1}$. Typical concentration of the residual orthohydrogen is 100 ppm. Dependence of the TPE intensity on the orthohydrogen concentration is measured using the targets with enriched orthohydrogen. The concentration is adjusted by mixing desired amount of normal-hydrogen gas into purified parahydrogen gas before the crystallization.

### B. Energy scheme and laser systems.

The experimental apparatus and energy scheme are summarized in Fig. 1. Two laser systems, a driving laser and a trigger laser, are used in the present study. The details of the laser systems are described in the previous papers.[19, 33] The $v = 1$ state of the parahydrogen is coherently excited by the stimulated Raman scattering with visible pulses from the driving laser system. It emits 532-nm pulses and 683-nm pulses simultaneously. The energy difference of the two pulses is about 4150 cm$^{-1}$, corresponding to the vibrational Raman transition energy of solid parahydrogen. The wavelength of the 683-nm pulses can be shifted by several GHz and it is possible to control two-photon detuning δ from the resonance. The detuning δ is calculated from the frequencies of the two driving pulses measured by a wavemeter (HighFinesse WS-7). The 532-nm pulses are the second harmonics of a Nd:YAG laser (Litron LPY642). The 683-nm pulses are prepared by optical parametric generation and amplification pumped by the identical Nd:YAG laser. These pulses are originated from the same oscillator so that time jitter between them is negligible. The two pulses are assumed to be coincided perfectly below. The pulse durations (full width at half maximum, FWHM) are about 9 ns for 532 nm and 6 ns for 683 nm, respectively. Previous studies have shown that narrow frequency widths of the driving pulses are essential for generating high coherence. Therefore, the Nd:YAG laser and the optical parametric processes are injection seeded with narrow linewidth continuous wave lasers. Injection seeding narrows frequency widths to 100 ~ 200 MHz. Although pulse energy of each driving pulse can be above 5 mJ/pulse, typical driving energy is set below 1 mJ/pulse to avoid damage of the target. Effect of the damage is discussed in the next section.

The two-photon emission is induced by mid-infrared laser pulses emitted from the trigger laser system. The system consists of an optical parametric generator, an amplifier, and a different frequency generator pumped by another Nd:YAG laser (Continuum, Surelite I). The system is also injection seeded. Typical pulse energy and pulse duration of the trigger field are 100 µJ/pulse and 2 ns, respectively. The wavelength of the trigger pulses is set to 4586 nm, which is near a half of the vibrational transition energy of solid parahydrogen.

### C. Detection of TPE.

The driving and the trigger pulses are aligned collinearly by dichroic mirrors and injected into the target. Diameters (D4σ) of the driving pulses at the target cell are adjusted to be 2.5 ~ 3.0 mm. When TPE is triggered by the 4586-nm photons, another 4586-nm photon and 5077-nm photon are emitted. The sum of the energies of these two photons is equal to vibrational energy of solid parahydrogen. Therefore, TPE can be observed by detecting the 5077-nm photons. The 5077-nm TPE signal is separated from incident laser pulses and the Raman scattering by frequency filters or a monochromator (Princeton Instruments, Acton SpectraPro SP2300). Then it is detected by a Mercury-Cadmium-Telluride (MCT) mid-infrared detector (Vigo systems, PC-3TE-9). Mutual timing between the driving pulses and the trigger pulse can be controlled by a delay generator (Stanford Research DG-645). In the present experiment, the trigger field also generates anti-Stokes scattering at 1580 nm simultaneously (Fig. 1 (a)). It corresponds to usual coherent anti-Stokes Raman scattering (CARS). CARS intensity is also measured for comparison.

## III. RESULTS AND DISCUSSIONS

### A. Observed spectrum.

Figure 2 shows a spectrum of one of the pair photons emitted via TPE, which is measured using the monochromator and the MCT detector. The peak wavelength and FWHM are determined to be 5077 nm and 3 nm, respectively, by the least-squares fitting with a Gaussian function (a solid line). The peak frequency agrees with the value expected from the energy conservation. The FWHM is limited by a resolution of the monochromator. Intensity of the 5077-nm signal depends on various experimental conditions such as driving pulse energy, target quality, and mutual timing between the driving and the trigger fields. Typical output energy of the 5077-nm signal is roughly estimated by considering transmittance of optical elements and sensitivity of the detector. It is a few nJ/pulse, corresponding to $10^{10} \sim 10^{11}$ photons/pulse. It is comparable to that of the previous gas experiments triggered by external fields.[19] The present target length is one thirtieth of that in the gas experiment. It suggests that efficiency of TPE per volume in solid is larger than that in gas phase. The high efficiency in solid is a consequence of coexistence of the high density and the long decoherence time. The excitation, development of the coherence, and TPE should have spatial structures along the direction of propagation. It is essential to investigate these structures or estimate effective target length for quantitative discussion of the efficiency. Further experiments and numerical simulations are desired. The present TPE intensity in solid is limited mainly by damage of solid due to the driving pulses in the current setup.

**B. Dependence of TPE intensity on the two-photon frequency detuning.**

Figure 3 shows typical dependence of the TPE intensity on the two-photon detuning δ. The driving and the trigger pulses are injected simultaneously (zero mutual timing). Pulse energy of each driving pulse is below 1 mJ/pulse. The zero of the horizontal axis indicates the resonance of the Raman transition. This resonance frequency is determined by the stimulated Raman gain spectroscopy with weak 532-nm pulses (about 20 µJ/pulse) and continuous wave 683-nm laser (about 1 mW). The resultant resonance energy is 4149.666(5) cm$^{-1}$. Error value is mainly due to uncertainty of the wavemeter. The transition frequency was reported to be 4149.6 ~ 4149.8 cm$^{-1}$ by several authors.[25, 34-36] Katsuragawa and Hakuta measured it carefully and reported it to be 4149.641(5) cm$^{-1}$.[36] Our measurement deviates from their result by 0.025 cm$^{-1}$ (750 MHz), which is five times as large as the error. They also reported that the transition frequency depends on pressures under which targets are prepared.[26] The transition frequencies at 24, 28, and 32 atm are 4149.654, 4149.648, and 4149.642 cm$^{-1}$, respectively. Our targets are prepared under much less than 1 atm so that the frequency discrepancy is attributed to the pressure difference during preparation. Although the transition frequency also depends on target temperature and residual orthohydrogen concentration, we define δ = 0 as 4149.666 cm$^{-1}$ in this paper. Peak detuning and FWHM of the detuning dependence curve are determined by fitting with a Gaussian function. The peak detuning averaged over seven data sets is -85 MHz. Considering the uncertainty, we conclude that the deviation from the resonance (δ = 0) is not significant under the present experimental conditions, although there are experimental and theoretical studies claiming that the maximum of the coherence is achieved at off-resonance.[37-41] The FWHM is in the order of 100 MHz and much larger than the reported Raman linewidth (~ 10 MHz). This is attributed to the linewidth of the driving fields (100~200 MHz). The detuning dependence of CARS intensity is also measured and it is found that its peak frequency and FWHM agree with those of TPE within experimental uncertainty. We also measure detuning dependence at different mutual timing and confirm that the peak frequency and FWHM do not depend on mutual timing within the uncertainty.

**C. Dependence on mutual timing between the driving and the trigger fields.**

Figure 4 shows typical dependences of TPE intensity on mutual timing before (open circles) and after annealing (closed circles) compared with dependence of the simultaneously generated CARS (cross marks). We define the mutual timing $t$ as the arrival time of the peak of the trigger field with reference to those of the driving fields. $t = 0$ means that peak of the trigger field coincides with those of the driving fields. $t > 0$ means that the trigger field reaches the target after the driving fields. All traces in Fig. 4 are normalized by each signal intensity at $t = 0$. The detuning $\delta$ is fixed at zero. Pulse energy of each driving field is below 1 mJ/pulse. It is clearly shown that TPE occurs even after the driving fields pass thorugh the target. This behavior is quite different from common four-wave mixing, which does not use real states and durable high coherence. The TPE intensity in the annealed target monotonically increases from about -5 ns to 5 ns where the driving fields still remain in the target (building-up region). Interestingly, TPE intensity develops without the driving fields up to 30 ns (developing region). Then, its intensity has maximum and starts to decay. The signal decays exponentially from about 80 ns to 300 ns (exponential decay region) and shows a non-exponential long tail after that. The CARS intensity shows similar temporal behavior to the TPE intensity. This is because that both TPE and CARS traces reflect the development of the coherence prepared by the driving fields. Time-resolved CARS of the $v = 1$ state[28] and the $v = 2$ state[29] of solid parahydrogen were reported previously. Their time scales of the decay are also in the order of 100 ns and agree with the present results. However, the decay curve of the previously reported time-resolved CARS deviates largely from an exponential curve, which is in contrast with near-exponential decay of this study. The decay curve is highly sensitive to the target properties: target annealing, damage induced by laser fields, target temperature and residual orthohydrogen concentration, as described below. The discrepancy of the decay curves is due to the different properties of solid parahydrogen. The non-exponential tail in the present experiments should reflect fine mechanisms of the decoherence similarly to the case of the time-resolved CARS.[28, 29] Further experiments with sufficient signal-to-noise ratio at the weak tails allow us to discuss the mechanism precisely.

The most interesting feature is the intensity development without driving fields from 5 ns to 30 ns (the developing region). It means the coherence of the vibrational state develops after the driving fields pass through the targets. This development is quite sensitive to the decoherence as shown below. Similar behaviors were also reported in the TPE in semiconductor[5] and the $v = 2$ CARS in solid parahydrogen.[29] Their origin is not understood well. It is confirmed that intensity of the driving fields at this region is at least two orders of magnitude smaller than the peak intensity. Therefore, it is difficult to attribute the development of the coherence to the tail of the driving fields. From the viewpoint of the simple optical-Bloch picture, this development seems very curious because it may be possible only with the inverted population and the present pump energy is not enough to make the inverted population. The origin of the development, therefore, is not included in the simple picture explicitly, for example, intermolecular interactions or interactions with environment. Although the origin is still unclear, several interesting phenomena, which may be related to the present behavior, have been discussed. They include intermolecular effects on the vibrational coherence[42], synchronization of quantum oscillators[43, 44], and interaction via coherent phonons.[45-47] Dedicated theoretical studies on such kind of intermolecular interactions are urged.

**C. 1 Effect of target annealing.**

It is clearly seen in Fig. 4 that the annealing drastically enhances durability of the coherence. Similar annealing effect was reported in previous study on time-resolved CARS of the $v = 2$ state of solid parahydrogen.[29] It is attributed to suppression of the coherence decay caused by elastic scattering of vibrational quanta (vibrons) with structural inhomogeneity, which is decreased by the annealing. This is the case in the present study. Annealed targets are used in all experiments described below.

**C. 2 Effect of laser-induced damage.**

It is found that the mutual timing dependence gradually changes during measurement when pulse energies of the driving fields are high. Four traces in Fig. 5 show change of the timing dependence at 4.1 K. The traces are measured successively. Energies of the 532-nm and 683-nm pulses are both about 4 mJ/pulse. The traces are normalized by the intensity of the first trace at $t = 0$. While the first trace shows similar behavior to that in Fig. 4, the decays of the latter traces become faster and deviate from exponential curve. The development after the driving fields also diminishes. It is also found that this change is recovered by the annealing above 10 K. After the annealing, the timing dependence is again almost the same as the first trace. This change of the mutual timing dependence is suppressed, even if it does not disappear, with lower driving fields energy, especially below 1 mJ/pulse. Considering these results, the change of the decay is attributed to damage of the target induced by the driving fields. The damage causes structural inhomogeneity in targets and shortens the decay time by the elastic scattering of the vibrons. The annealing eliminates such defects and recover the decay time. It is worth noting that signal intensities at $t = 0$ do not drastically change. This suggests that the developing region is more sensitive to the decoherence than the building-up region.

**C. 3 Effect of target temperature.**

Figure 6 shows temperature effect on the TPE intensity. The upper panel and lower panel show the detuning dependence at $t = 0$ and the timing dependence for various temperatures. The timing dependence curves are measured at the detuning where TPE intensity is maximum at each temperature. The traces are not normalized. At higher temperature, (1) the detuning traces are shifted to negative side, peak intensity becomes lower and linewidth becomes broader, (2) the peak intensities of the timing curves become lower and their decays become faster, (3) duration of the developing region becomes shorter and its effect becomes smaller, (4) timing of maximum intensity becomes earlier. For more quantitative discussions, the detuning curves are fitted by a Gaussian function, and their peak intensities, FWHMs, and peak detunings are determined. The timing curves around their maximums are also fitted by a Gaussian function to evaluate their peak intensities and peak timings. Their exponential decay regions are separately fitted with an exponential decay function: $A + B \exp(-2t/\tau)$. Fitting results are summarized in Fig. 7.

Figure 7 (a) shows the peak detunings of the detuning curves. Shift of the resonance frequency of the vibrational Raman transition of solid parahydrogen by temperature has been studied previously.[26, 48] It is explained by elastic scattering of the vibrons with two phonons whose distribution strongly depends on temperature. The temperature dependence of the shift is represented as $-(\pi^4/15)\alpha(T/T_D)^4$ in the low temperature limit, where $T$ is temperature and $T_D$ is the Debye temperature of solid parahydrogen. The $\alpha$ is a constant representing the coupling strength between the vibrational motion of hydrogen molecules and the phonon. The Debye temperature of solid parahydrogen is about 100 K.[26, 48, 49] In the present paper, $T_D = 108$ K is assumed according to the reference 48. Solid line in Fig. 7 (a) shows the fitted line with a function $v_0 - (\pi^4/15)\alpha(T/T_D)^4$. The $v_0$ is peak detuning in the limit of $T = 0$ K. The observed data are well represented by the function. It suggests that the shift of the detuning curves can be attributed to the scattering by the phonons. The $\alpha$ is determined to be 1.68(1) THz, which slightly deviates from the reported value of 1.8 THz.[48] The error values in this subsection reflect only the statistical uncertainties.

Figure 7 (b) shows FWHMs of the detuning curves and $1/\pi\tau$ obtained from the timing curves. The decay constant $\tau$ can be regarded as transverse relaxation time so that $1/\pi\tau$ corresponds to FWHM of the spectra. Although both values in the Fig. 7 (b) increase monotonically, the widths determined by the decay constants are an order of magnitude narrower than those determined by the detuning curves. This is probably due to laser linewidth. An offset component of the FWHM of the detuning curves is about 200 MHz, which agrees with the linewidths of the driving fields. In other words, the decay of the coherence is not directly affected by the laser linewidth. The linewidth of the vibrational Raman transition of solid parahydrogen has also been discussed in the view point of the two-phonon scattering. The temperature dependence of FWHM of the Raman transition is represented as $2(16\pi^6/21)\beta(T/T_D)^7$ at low temperature.[26, 29, 48] The $\beta$ is a constant representing the coupling strength between phonons and vibrons. A solid line in Fig. 7 (b) shows fitting result to the observed data with a function $2\Gamma_0 + 2(16\pi^6/21)\beta(T/T_D)^7$. The $2\Gamma_0$ represents offset linewidth which remains in the limit of $T = 0$ K. The $\Gamma_0$ and $\beta$ are determined to be 2.91(3) MHz and 321 (12) GHz, respectively. These values also deviate slightly from the reported value, 4.2 MHz and 390 GHz.[48] Considering different properties of the target due to the different preparation conditions, small disagreements of $\alpha$, $\beta$, and $\Gamma_0$ are acceptable.

Figure 7 (c) shows the maximum TPE intensity of the detuning curves at $t = 0$ (closed circles) and that of the timing curves at the peak detuning (open circles). Each intensity is normalized with that at $T = 4$ K. While both data show similar monotonically decreasing behaviors, decay of the timing curve is slightly faster than that of the detuning curves. Decrease of intensity is probably due to the fast decoherence at higher temperature. Difference between the two curves comes from different sensitivities to the decoherence. As is pointed out in the discussion of the effect of the laser induced damage, the coherence development after the driving field is more sensitive to the decoherence than the coherence building up by the driving fields. Figure 7 (d) shows mutual timings at the peak of the timing curve. It becomes earlier at higher temperature. This also suggests that the developing region is sensitive to the decoherence.

### C. 4 Effect of orthohydrogen concentration.

Figure 8 shows the TPE intensity at various orthohydrogen concentrations. The upper panel shows the detuning curves at $t = 0$. TPE intensity becomes weaker at higher concentration so that pulse energies of the driving fields are adjusted in order to obtain enough signal-to-noise ratio. Therefore, we do not discuss the intensity change further and all traces are normalized by their maximum values. It is clearly seen that the traces shift to negative and the FWHM becomes larger at higher concentrations. The peak detuning and the FWHM are determined by fitting with a Gaussian function (solid lines). These values are summarized in Fig. 9. The lower panel in Fig. 8 shows the timing dependence, which is also normalized by the intensity at $t = 0$. The timing dependence curves are measured at detuning where the TPE intensity is maximum at each concentration. The orthohydrogen concentration dependence of the timing curves is very similar to the temperature dependence. This is because decohenrece occurs due to elastic scattering not only with phonons but also with impurities, as discussed in previous paper.[29] Decay constants and peak mutual timings are determined by the same method in the case of the temperature dependence. They are also summarized in Fig. 9.

Figure 9 (a) shows peak detunings of the detuning curves (closed circles). They show linear dependence on orthohydrogen concentration. The slope is determined by fitting with a linear function to be -941.6(5) MHz/%. The frequency shift of the vibrational Raman transition of solid hydrogen was studied both experimentally and theoretically.[22] The origin of the shift is reduction of vibron hopping interaction by impurities, that is to say, resultant shrinkage of the vibron band. The theoretical value of the slope of solid parahydrogen at zero pressure is -882 MHz/%.[22] It agrees with the present work except for a little difference, which is attributed to imperfection of the actual crystal. It suggests that peak detunings coincide with the resonance frequency even at higher orthohydrogen concentration. In Fig. 9 (a), the present data are compared with the experimental Raman transition energy reported previously.[50] Both slopes are in good agreement while there are offset of about 1 GHz. As described above, the absolute transition frequency is sensitive to not only orthohydrogen concentration but also other properties such as pressure under which targets are prepared.

Figure 9 (b) shows FWHMs and corresponding decay constants. The FWHMs of the detuning curves are an order of magnitude larger than those calculated from the decay constants same as the case of the temperature dependence. FWHMs determined by the decay constants in the present study are narrower than the reported linewidth of the stimulated Raman spectroscopy although qualitative behavior is similar.[50] The FWHM by the decay seems to be saturated above 50000 ppm in contrast to that of the detuning curve. This is because that the timing curves at higher orhohydrogen concentration are determined by the temporal lineshape of the driving fields. The timing curves above 50000 ppm are compared with the lineshapes of the driving fields in Fig. 10. It is clearly seen that the timing curves at 50000 ppm and 120000 ppm are the same. The building-up of the TPE intensity is slightly slower than those of the driving fields. The decays seem to be determined by the driving fields. This suggests the insensitivity of the building-up region to the decoherence. Figure 9 (c) shows the mutual timing at the peak of timing curves. It decays rapidly with orthohydrogen concentration and seems to be saturated above 50000 ppm. This rapid decay also suggests that the developing region is sensitive to the decoherence.

## IV. CONCLUSIONS

Two-photon emission from the $v = 1$ state of solid parahydrogen coherently excited by the Raman scattering is observed. The emission is triggered by the mid-infrared pulses. Intensity of TPE is maximum when the detuning of the driving fields is near the two-photon resonance. The TPE is observed when the trigger pulse is injected several hundred nanoseconds after the driving fields pass through the targets. The mutual timing dependence of TPE is similar to that of time-resolved CARS with a long tail, which is due to usage of the real state and the long decoherence time of solid parahydrogen. Interestingly, the coherence develops even after the driving fields pass through the targets. It is also found that this development is quite sensitive to the decoherence. Although further experimental and theoretical studies are needed to understand the phenomena, it is probably caused via intermolecular interactions. The decoherence is enhanced by the laser-induced damage on the target, high target temperature and high residual orthohydrogen concentration while it is suppressed by the target annealing. These effect can be understood by elastic scattering of vibrons by thermal phonons or inhomogeneity.

The present TPE from solid parahydrogen provide new attracting light sources. It can be used as a frequency conversion technique because the energy of emitted photons is equal to energy difference between the vibrational state and the trigger pulse. There are several reports on frequency conversion using solid parahydrogen as Raman shifter media.[51-53] It is, however, difficult to generate far- infrared light because strong absorption of solid parahydrogen above 4150 cm$^{-1}$. The TPE is free from absorption of solid parahydrogen because there is no absorption below 4150 cm$^{-1}$ except for several rotational absorption bands. The two-photon emission is also interesting as two-photon light source which generates correlated photons. It is possible to generate photon pair at arbitrary timing within the decoherence time.

## ACKNOWLEDGMENTS

YM thanks Dr. S. Kuma and Dr. T. Hiraki for useful discussions and suggestions. This research was partially supported by Grant-in-Aid for Scientific Research A (15H02093), Grant-in-Aid for Challenging Exploratory Research (15K13486), Grant-in-Aid for Young Scientists B (15K17651) from the Ministry of Education, Culture, Sports, Science, and Technology.

[48] F. L. Kien, A. Koreeda, K. Kuroda, M. Suzuki, K. Hakuta, *Jpn. J. Appl. Phys.* **42**, 3483 (2003).

[49] P. C. Souers, *Hydrogen Properties for Fusion Energy*; University of California Press: Berkeley, (1986).

[50] T. Momose, T. Oka, *J. Low Temp. Phys.* **139**, 515 (2005).

[51] B. J. McCall, A. J. Huneycutt, R. J. Saykally, C. M. Lindsay, T. Oka, M. Fushitani, Y. Miyamoto, T. Momose, *Appl. Phys. Lett.* **82**, 1350 (2003).

[52] M. Fushitani, S. Kuma, Y. Miyamoto, H. Katsuki, T. Wakabayashi, T. Momose, *Opt. Lett.* **28**, 37 (2003).

[53] K. E. Kuyanov, T. Momose, A. E. Vilesov, *Appl. Opt.* **43**, 6023 (2004).
13

**FIGURES**

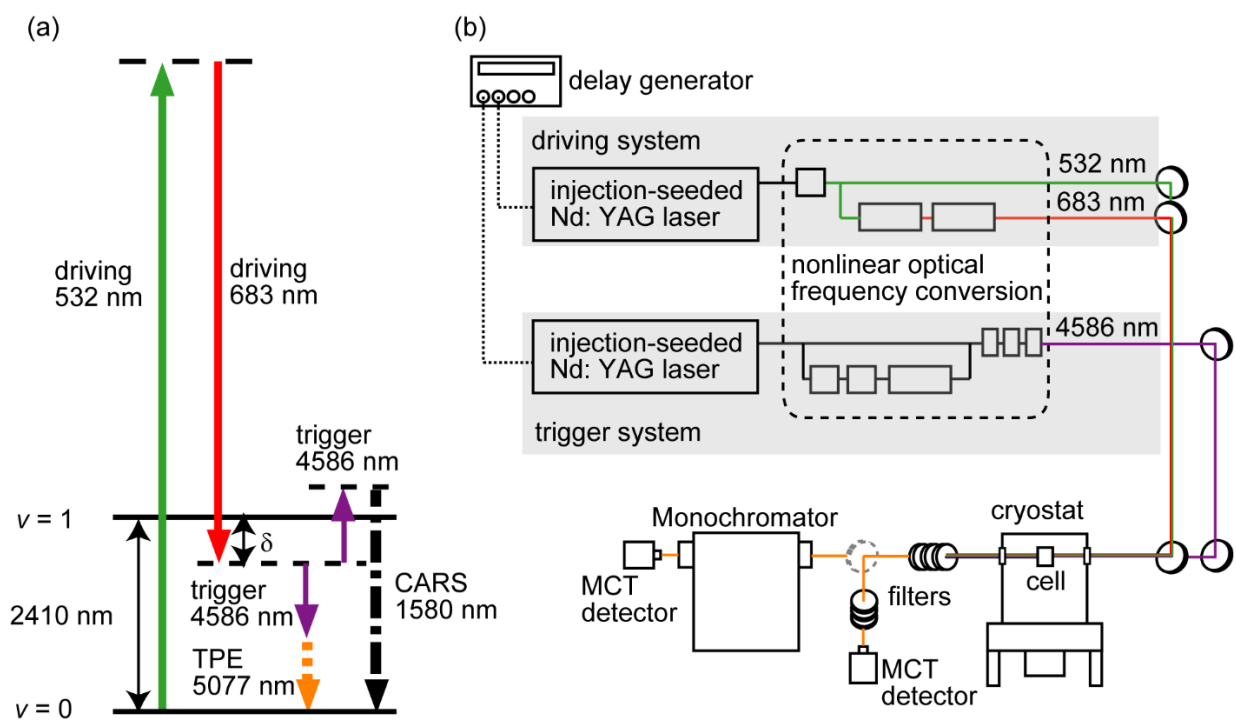

FIG. 1. (a) Energy diagram and (b) experimental setup.

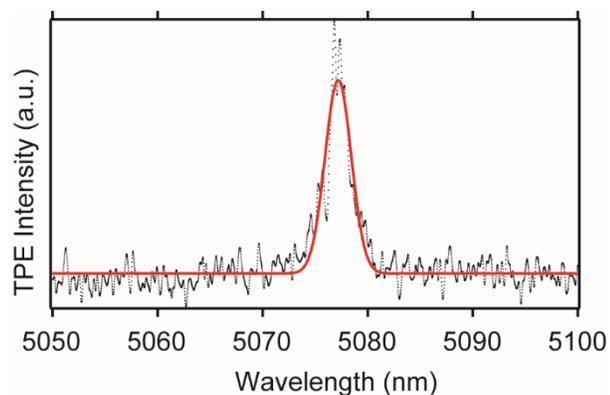

FIG. 2. Observed spectrum of TPE. A solid red line is a fitting result with a Gaussian function.

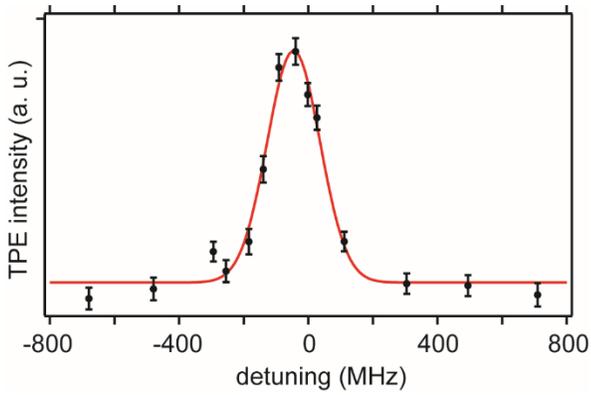

FIG. 3. Dependence of TPE intensity on two-photon detuning of the driving fields. The error bars indicate the standard errors. A solid red line is a fitting result with a Gaussian function.

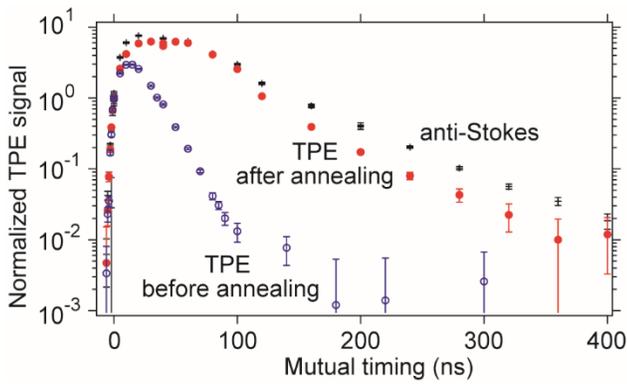

FIG. 4. Dependence of TPE intensity on mutual timing between the driving fields and the trigger field before (open blue circles) and after annealing (closed red circles). Cross marks show intensity of anti-Stokes scattering generated simultaneously by the trigger pulses. The error bars indicate the standard errors.

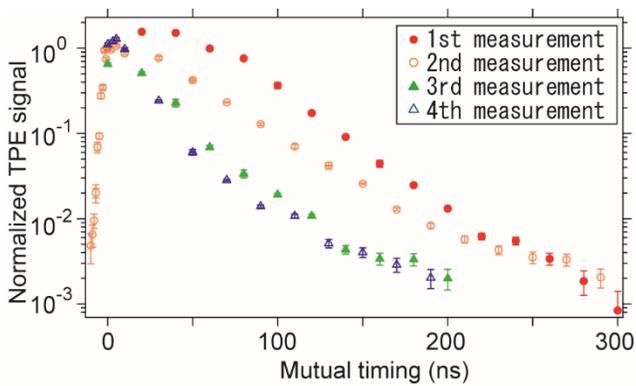

FIG. 5. Dependence of intensity of TPE on mutual timing observed successively. Closed red circles, open orange circles, closed green triangles and open blue triangles show the first, second, third and fourth measurement. The error bars indicate the standard errors.

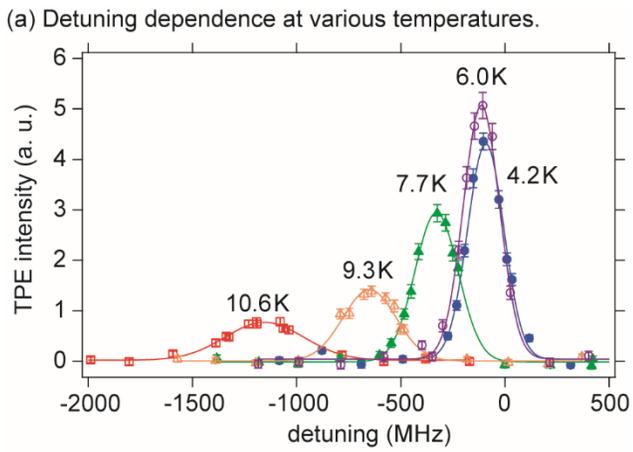

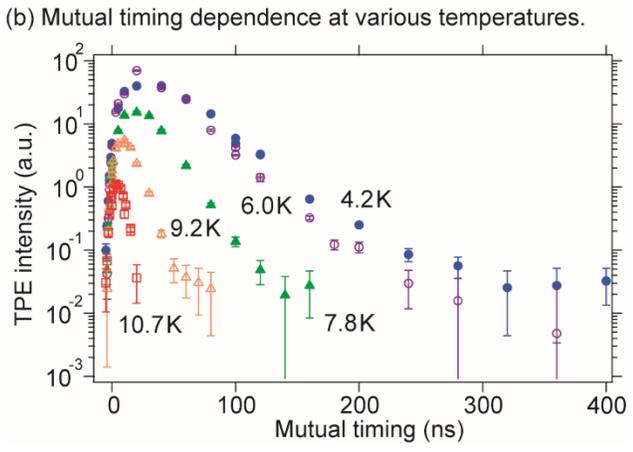

FIG. 6. (a) Detuning dependence of intensity of TPE at various temperatures. (b) Mutual timing dependence of intensity of TPE at various temperatures. The error bars indicate the standard errors.

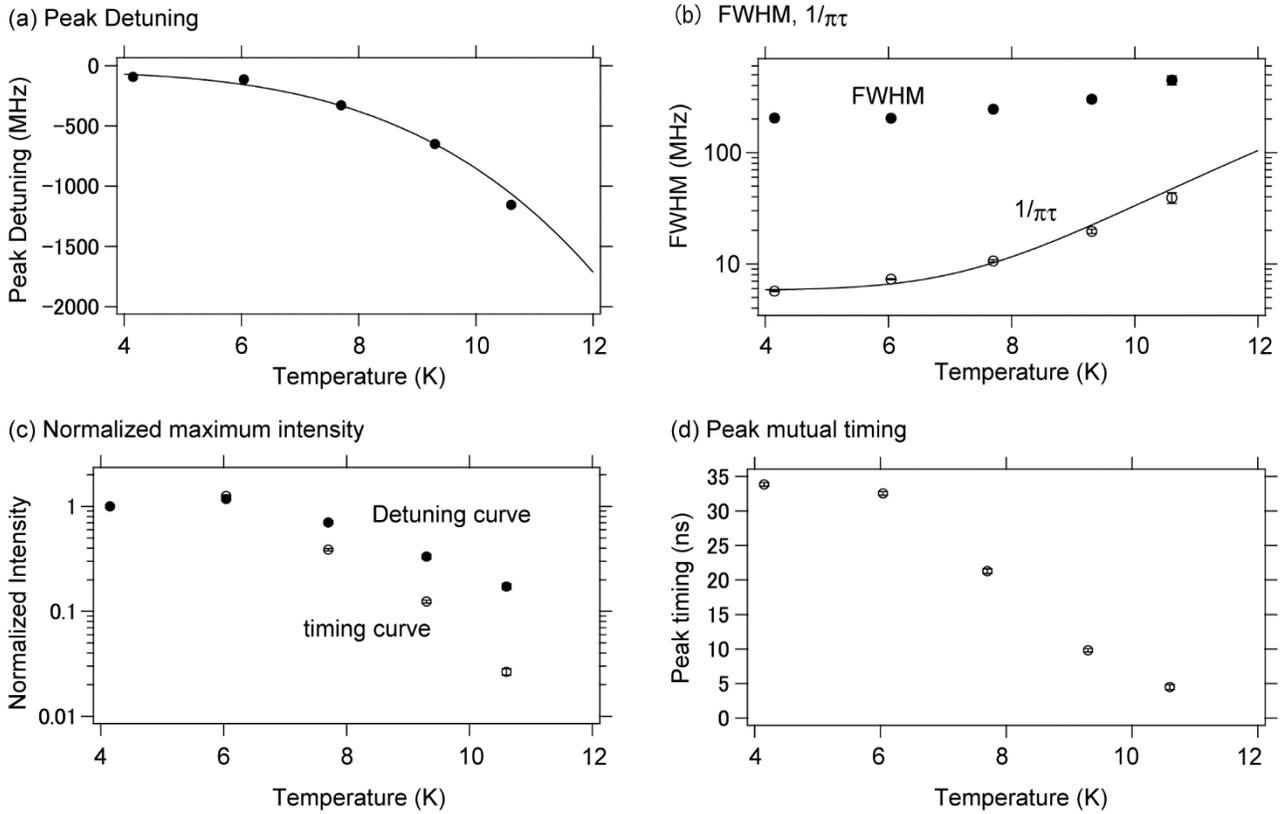

FIG. 7. Fitting results of temperature dependence of TPE. (a) Detuning at peak of detuning curves. (b) FWHM of detuning curves (closed circles) and decay constants of timing curves (open circles). (c) Maximum intensities of detuning curves (closed circles) and timing curves (open circles). Both data are normalized by the intensity at 4 K. (d) Mutual timing at peak of timing curves. The error bars indicate the standard errors.

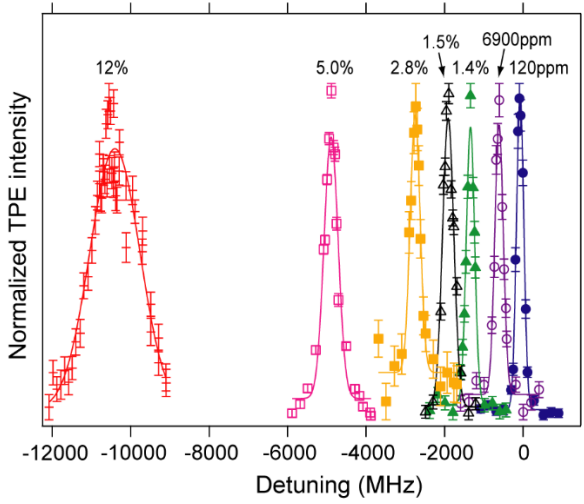

(a) Detuning dependence at various orthohydrogen concentrations.

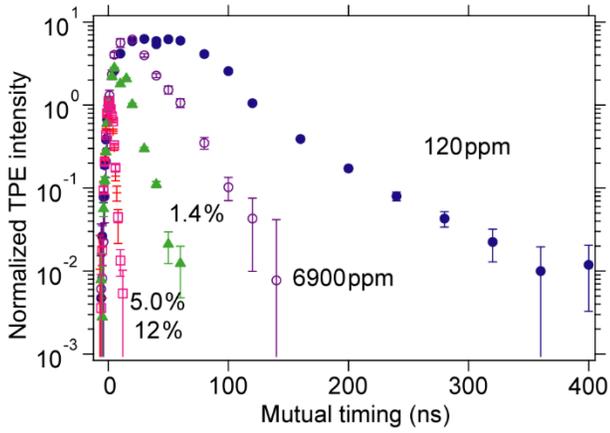

(b) Mutual timing dependence at various orthohydrogen concentrations.

FIG. 8. (a) Detuning dependence of intensity of TPE at various orthohydrogen concentrations normalized by each peak intensity. (b) Mutual timing dependence of intensity of TPE at various orthohydrogen concentrations normalized by intensity at $t = 0$.

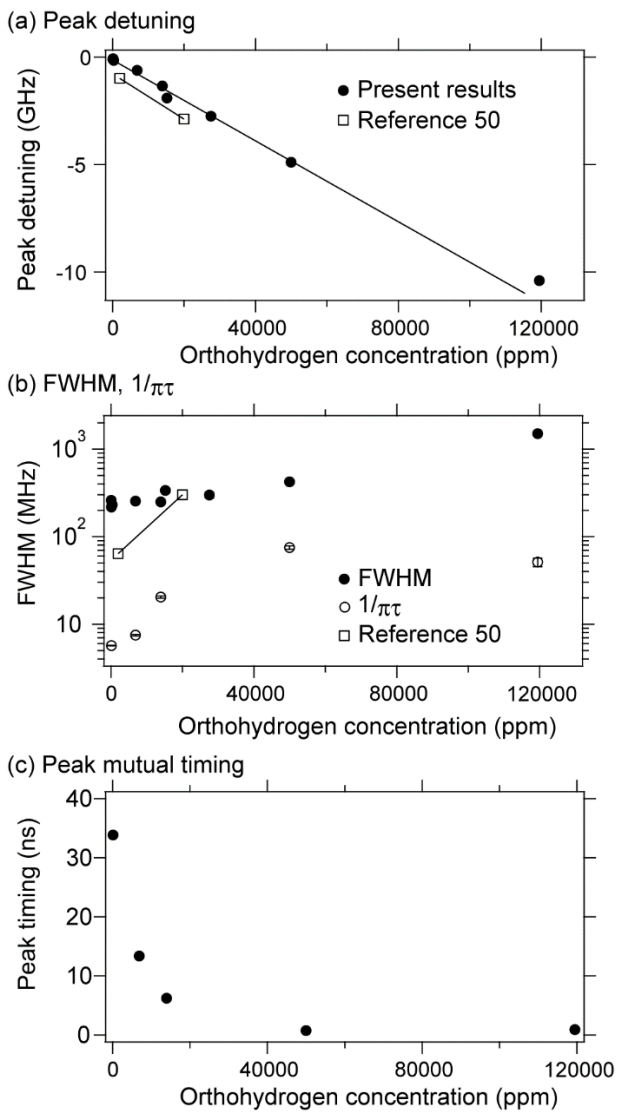

FIG. 9. Fitting results of orthohydrogen concentration dependence of TPE. Open squares show the data reported by the reference 50. (a) Detuning at peak of detuning curves. (b) FWHM of detuning curves (closed circles) and decay constants of timing curves (open circles). (c) Mutual timing at peak of timing curves. The error bars indicate the standard errors.

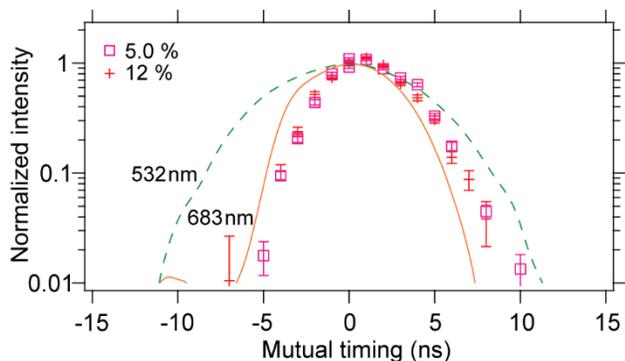

FIG. 10. Timing curves with orthohydrogen concentration of 5.0 % (open pink squares) and 12 % (red cross marks). The error bars indicate the standard errors. A solid orange line and a dashed green line show temporal pulse shapes of 683-nm and 532-nm pulses, respectively, which are averaged over 100 shots.